\documentclass[prd,preprintnumbers,nofootinbib]{revtex4} 
\usepackage{amsmath}
\usepackage{amssymb}
\usepackage{latexsym}
\usepackage{amsfonts}
\usepackage{slashed}
\usepackage{color}
\usepackage{graphicx}

\usepackage{hyperref}

\def\hhref#1{\href{http://arxiv.org/abs/#1}{arXiv:#1}}

\newcommand{\mc}{\mathcal}
\newcommand{\del}{\partial}
\newcommand{\bea}{\begin{eqnarray}}
\newcommand{\ea}{\end{eqnarray}}
\newcommand{\eea}{\end{eqnarray}}
\newcommand{\nn}{\nonumber\\}
\newcommand{\ord}{\,{\cal O}}
\newcommand{\B}{{\bf{\mathcal B}}}

\begin{document}
\title{The Chern-Simons diffusion rate in strongly coupled ${\cal N}=4$ SYM plasma \\ in an external magnetic field} 
\author{G\"ok\c ce Ba\c sar$^a$, Dmitri E. Kharzeev$^{a,b}$}
\affiliation{
$^a$ Department of Physics and Astronomy, Stony Brook University, Stony Brook, NY 11794, USA\\
$^b$ Department of Physics, Brookhaven National
  Laboratory, Upton, NY 11973, USA
}
\date{\today}

\begin{abstract}
We calculate the Chern-Simons diffusion rate in a strongly coupled $\mc N=4$ SUSY Yang-Mills plasma in the presence of a constant external $U(1)_R$ magnetic flux via the holographic correspondence. Due to the strong interactions between the charged fields and non-Abelian gauge fields, the external Abelian magnetic field affects the thermal Yang-Mills dynamics and increases the diffusion rate, regardless of its strength. We obtain the analytic results for the Chern-Simons diffusion rate both in the weak and strong magnetic field limits. In the latter limit, we show that the diffusion rate scales as $B\times T^2$ and this can be understood as a result of a dynamical dimensional reduction. 
\end{abstract}

\maketitle

\section{Introduction}

 The non-Abelian gauge fields possess a rich topological structure. The vacuum contains an infinite number of energy-degenerate sectors characterized by an integer: the Chern-Simons number $N_{CS}$ that is a topological quantity determined by the global structure of the gauge fields. 
 At zero temperature, the different Chern-Simons sectors are connected by quantum tunneling transitions (the instantons \cite{Belavin:1975fg,'tHooft:1976fv}); this leads to the picture of ``$\theta$- vacuum'' of non-Abelian gauge theories. 
 At finite temperature, the gauge configurations that change the Chern-Simons number can also be activated thermally. We will, loosely speaking, refer to these configurations as ``sphalerons" both at weak and at strong coupling, even though this term is usually reserved for describing the classical solutions at weak coupling \cite{Manton:1983nd,Klinkhamer:1984di}. As opposed to the tunneling processes (instantons), the sphaleron rate is not necessarily exponentially suppressed \cite{Kuzmin:1985mm,Arnold:1987mh}. 
 \vskip0.3cm
 The change of Chern-Simons number in such a process is given in terms of the topological Pontryagin invariant:
 
\bea
\Delta N_{CS}=\frac{g^2}{32 \pi^2} \int d^4x\ F^a_{\mu\nu} \tilde F_a ^{\mu\nu} (x)=\frac{g^2}{8 \pi^2} \int d^4x \,\text{tr}\bf E \cdot \bf B ,
\ea
where $\bf E$ and $\bf B$ are non-Abelian electric and magnetic fields, and $g$ is the Yang-Mills coupling.  The rate of change of Chern-Simons number is called the Chern-Simons diffusion rate
$\Gamma_{CS}$. It is simply the probability of a Chern-Simons number changing process to occur per unit volume and per unit time:

\bea
\Gamma_{CS}=\frac{\langle\Delta N_{CS}^2\rangle}{V\,t}=\int d^4x  \left\langle \frac{g^2}{32 \pi^2}F^a_{\mu\nu} \tilde F_a ^{\mu\nu} (x) \frac{g^2}{32 \pi^2}F^a_{\alpha\beta} \tilde F_a ^{\alpha\beta}(0)\right\rangle
\label{rate_def}
\ea
\vskip0.3cm
The weak coupling result for the diffusion rate for $SU(2)$ gauge theory is given by \cite{Bodeker:1998hm}:
\bea
\Gamma_{CS}&=&\kappa^\prime g^{10} \log(1/g^2) T^4\qquad\text{ (SU(2), weak coupling) },
\ea
where $\kappa^\prime$ is a (numerically large) constant.
This expression can be understood as a result of the Langevin--type dynamics of the non-Abelian gauge fields \cite{Bodeker:1998hm,Moore}.  Since the sphaleron at the peak of the barrier separating Chern-Simons sectors is a purely magnetic field configuration, the factor $(g^2 T)^3$ in this expression can be understood as the inverse magnetic screening length that determines the characteristic inverse volume of the sphaleron, and $g^4 T \log(1/g)$ -- as the typical inverse time scale of the process. Numerical calculations show that the constant $\kappa^\prime\sim10$ \cite{Moore,Moore2}  for $SU(2)$. 
\vskip0.3cm
The holographic AdS/CFT correspondence \cite{Maldacena,Gubser,Witten} makes it possible to compute the Chern-Simons diffusion rate in $\mc N=4$ SYM plasma at $N\rightarrow\infty$ in the strong coupling regime \cite{Son:2002sd}. 
The strong coupling result is given by \cite{Son:2002sd}:
\bea
\Gamma_{CS}&=&\frac{(g^2 N)^2}{256 \pi^3} T^4\qquad\text{ ($\mc N=4$ SYM, strong coupling, large N) }
\label{strongcoupling}
\ea 
It is remarkable that compared to the weak coupling case, at strong coupling the diffusion rate is substantially enhanced. This means that Chern-Simons diffusion is not exclusively the property of semi-classical fields, as the (Minkowski boundary) dynamics at strong coupling has to be driven by quantum effects.
\vskip0.3cm
The Atiyah-Singer index theorem relates the change in $N_{CS}$ to the chirality change in the fermionic sector -- every Chern-Simons number changing transition is accompanied by the flip of chirality. In thermal equilibrium, the sphalerons thus lead to the decay of any excess chiral charge $N_5=\langle J^0_5\rangle$ present in the medium. A linear response relation for a small chiral chemical potential leads to \cite{Moore, Moore2} the decay:  
\bea
\frac{d N_{5}}{dt}=-C\,N_5\frac{\Gamma_{CS}}{T^3}
\ea
where the constant $C$ depends on the details of the fermionic sector.
\vskip0.3cm
These $N_{CS}$-changing transitions violate locally P and CP symmetries and lead to very interesting consequences in a number of various physical settings. In weak interactions, they are related to the violation of baryon plus lepton $(B+L)$ number\footnote{We will refer to the violation of $(B+L)$ simply as baryon number violation from now on.} and play a crucial role in electroweak baryogenesis scenarios \cite{Kuzmin:1985mm,Arnold:1987mh,baryo1,baryo2,baryo3,baryo4,McLerran:1990de,Rubakov:1996vz}. In strong interactions, there exists a very strong constraint on the amount of global P and CP violation; this constraint originates  mainly from the experimental upper bound for the neutron dipole moment \cite{Baker:2006ts}. However on theoretical grounds, the observed CP invariance is still lacking a conclusive explanation (``the strong CP problem"). On the other hand, the instantons and sphalerons provide explicit examples of the \textit{local} fluctuations of topological charge leading, through the index theorem, to the local imbalance of chirality in QCD plasma. In a medium with high enough temperature, the fluctuations might lead to observable effects such as the ``chiral magnetic effect" \cite{CME} which is an induction of an electric current by an external magnetic field in the presence of Chern-Simons number changing processes. 
\vskip0.3cm
In this paper we address the effect that an external Abelian magnetic field has on the topological fluctuations in the plasma. This question has been addressed in the past in several different contexts. The chiral magnetic effect is seen in various lattice simulations \cite{ITEP,yamamoto,tom}.  The effects of magnetic field on an instanton configuration is also studied on the lattice \cite{ITEP2,tom2}. In \cite{Basar:2011by} the Dirac spectrum of an instanton in the presence of magnetic field is analyzed and it is argued that in strong magnetic field regime some of the physical quantities such as the magnetic and electric dipole moments are dominated by the (near-) zero modes. In \cite{Comelli:1999gt}, it is shown that the diffusion rate due to an electoweak sphaleron in the Higgs phase increases with a presence of an external magnetic field. 
\vskip0.3cm
Here we calculate the diffusion rate for strongly coupled $\mc N=4$ plasma in the presence of an external magnetic field via holography. The dual gravity is characterized by a full solution of asymptotically $AdS_5$, five-dimensional Einstein-Maxwell system with a constant magnetic flux.  Since the solution takes into account all of the back-reaction of the magnetic field on the metric, this gravity description allows one to work with arbitrarily strong magnetic fields. This configuration was previously studied in \cite{D'Hoker:2009mm} and, together with its various extensions, was pursued in a different context in \cite{D'Hoker:2011xw}.
\vskip0.3cm
The rest of the paper is organized as follows. In section \ref{sec_brane}, we summarize the magnetic brane solution presented in \cite{D'Hoker:2009mm} which constitutes our dual metric. Then, in section  \ref{sec_rate}, we calculate the diffusion rate and present the exact numerical result. Section \ref{sec_limits}, is devoted to the high temperature and low temperature limits of this result, for which we obtain the analytic expressions. At high temperature we calculate the leading term to the zero magnetic field result (\ref{strongcoupling}) analytically by treating the magnetic field perturbatively. At low temperature, we use the (2+1)-dimensional  BTZ black hole solution to describe the dimensionally reduced metric and express the rate also analytically at this limit. We conclude the paper by arguing that the effect is negligible for heavy ion collisions and discuss a possible implication for a particular scenario of electroweak baryogenesis.

\section{The dual geometry}
\label{sec_brane}

Let us briefly review the magnetic brane solution studied in \cite{D'Hoker:2009mm} that constitutes the gravity dual of our problem. The basic setup is the five-dimensional Einstein-Maxwell theory with a negative cosmological constant. In the boundary gauge theory, the Abelian Maxwell field  is associated with the $U(1)_R$ symmetry which the gauginos, the chiral multiplet fermions and scalars are charged under. The action also contains the five-dimensional Chern-Simons term that accounts for the $U(1)_R$ anomaly:
\bea
S=-\frac{1}{16\pi G_5}\int d^5x\sqrt{-g}\left(R+F^{MN}F_{MN}-\frac{12}{l^2}\right)+\frac{1}{6\sqrt{3}\pi G_5}\int A\wedge F\wedge F+S_{bdry}
\label{action}
\ea
The last term in (\ref{action}) is the boundary term that is determined by imposing a sensible variational principle. The AdS radius $l$ will be set to unity for the rest of the paper. It is also worth to mention that the coefficient of the variation of the Chern-Simons term is fixed by the $U(1)_R$ anomaly of the gauge theory. This relates the field strength $F$ in (\ref{action}) to the physical field strength in the gauge theory $\mc F$ as $\mc F= \sqrt{3} F$. We refer the reader to \cite{D'Hoker:2009mm} for further details.
\vskip0.3cm

Let us now introduce a constant magnetic flux in the $x_3$ direction:
\bea
F=B\ dx^1 \wedge dx^2
\label{flux}
\ea
The constant flux satisfies the Maxwell equations trivially. For the Einstein equations
\bea
R_{MN}=4g_{MN}+\frac{1}{3}F^{AB}F_{AB}g_{MN}-2F_{MA}F_N^{\,A}
\label{einstein}
\ea
we start with the general form of the 5-dimensional metric
\bea
ds^2=-U(r)dt^2+\frac{dr^2}{U(r)}+e^{2V(r)}(dx_1^2+dx_2^2)+e^{2W(r)}dx_3^2 ,
\label{metric}
\ea
and seek for asymptotically $AdS_5$ solutions with a horizon to study the problem at nonzero temperature, in accordance with the AdS/CFT correspondence. Unfortunately, there is no known analytical solution to the Einstein equations satisfying these properties. However it is possible to solve them numerically. Notice that the metric components $U, V$ and $W$ are functions of a single variable $r$. Therefore the Einstein equations (\ref{einstein}) are a set of coupled ordinary differential equations in $r$. Following \cite{D'Hoker:2009mm}, we rescale the coordinates such that the horizon is at $r=1$ (i.e $U(1)=0$), and $U^\prime(1)=1, V(1)=W(1)=0$. The remaining two initial conditions $V^\prime(1)$ and $W^\prime(1)$ are encoded in the Einstein equations. Then, given the initial conditions, we integrate the Einstein equations from the horizon $r=1$ to the boundary $r\rightarrow\infty$ numerically. The  result is that the solutions indeed are asymptotically $AdS_5$:
\bea
U(r)\rightarrow  r^2,\quad e^{2V(r)}\rightarrow v r^2,\quad e^{2W(r)}\rightarrow w r^2\quad \quad as \,\,r\rightarrow\infty
\ea
Here $v$ and $w$ are the functions of the magnetic field strength $B$. To get the correct $AdS_5$ limit on the boundary, we should rescale the $x_1,x_2$ and $x_3$ coordinates: 
\bea
ds^2=-U(r)dt^2+\frac{dr^2}{U(r)}+\frac{e^{2V(r)}}{v}(dx_1^2+dx_2^2)+\frac{e^{2W(r)}}{w}dx_3^2
\label{metric2}
\ea
This rescaling also affects the form of the physical magnetic flux:
\bea
F=\frac{B}{v}dx_1\wedge dx_2
\label{flux2}
\ea 
It is also useful to define the physical magnetic field strength in the gauge theory $\B$:
\bea
\B=\sqrt{3}\frac{B}{v}
\ea

\section{The Chern-Simons diffusion rate}
\label{sec_rate}

The diffusion rate (\ref{rate_def}) is nothing but the zero frequency and wavelength limit of the symmetrized Wightman correlator of the topological charge:
\bea
G^{sym}(\omega,\vec k)\equiv\frac{1}{2}\int d^4x e^{-ikx}\left\langle \left\{\frac{1}{4}F^a_{\mu\nu} \tilde F_a ^{\mu\nu}(x)\,,\,\frac{1}{4}F^a_{\mu\nu} \tilde F_a ^{\mu\nu}(0)\right\}\right\rangle
\ea
The fluctuation-dissipation theorem relates the symmetrized Wightman correlator to the imaginary part of the Green's function:
\bea
G^{sym}(\omega,\vec k)=-\text{coth}\,\left(\frac{\omega}{2 T}\right)\,\mc Im [G^R(\omega,\vec k)]\approx-\frac{2 T}{\omega}\,\mc Im[G^R(\omega,\vec k)]
\label{fdt}
\ea
where the retarded Green's function is:
\bea
G^R(\omega,\vec k)\equiv-i\int d^4x e^{-ikx}\theta(t)\left\langle \left[\frac{1}{4}F^a_{\mu\nu} \tilde F_a ^{\mu\nu}(x)\,,\,\frac{1}{4}F^a_{\mu\nu} \tilde F_a ^{\mu\nu}(0)\right]\right\rangle\\
\ea
Hence, to extract the diffusion rate all we need to know is the zero momentum and small frequency limit of the retarded Green's function:
\bea
\Gamma_{CS}&=&-\left(\frac{g^2}{8 \pi^2}\right)^2\, \lim_{\omega\rightarrow0}\frac{2 T}{\omega}\mc Im \left[G^R(\omega,\vec k=0)\right]
\label{rate_fdt}
\ea
The imaginary part of the retarded Green's function is associated with the thermal dissipation and therefore is a T-odd function. As a result the leading term in the small frequency expansion has to be linear in $\omega$, with the coefficient that controls the diffusion rate (\ref{rate_fdt}) .
\vskip0.3cm
We compute the retarded Green's function $G^R$ of the topological charge by using the AdS/CFT duality where the gravity side is described by the metric (\ref{metric2}). The duality relates the operator $\frac{1}{4}F^a_{\mu\nu} \tilde F_a ^{\mu\nu}$ in $\mc N=4$ SYM with the RR scalar $C_0$ which is the axion in the ten-dimensional type IIB supergravity theory. In the absence of the $U(1)_R$ Maxwell field, the axion is nothing but a free massless scalar in the consistent truncation to the five-dimensional effective theory after Kaluza-Klein reduction of the $S^5$. To the second order in fluctuations in the fields, it remains as a free scalar even with the presence of a background $U(1)_R$ flux. To see this let us analyze the axion part of the ten-dimensional type IIB SUGRA action in the string frame:
\bea
S_{axion}\propto\int\sqrt{-G}(|dC_0|^2+|C_2-C_0\wedge dB|^2) ,
\label{axion}
\ea 
where $C_0$ is the axion, $C_2$ and $B$ are anti-symmetric RR and NS-NS  2 forms respectively. In the gauge theory, the $U(1)_R$ symmetry is realized as the diagonal $U(1)_R$ of the Cartan subgroup $U(1)_R^3$ of $SU(4)_R$. In the SUGRA picture, this corresponds to wrapping around the 3 angles of the $S^5$ simultaneously by a special choice of truncation parameters. The Kaluza-Klein reduction can be written as \cite{Chamblin:1999tk,Cvetic:1999xp}: 
\bea
ds^2=g_{MN}dx^Mdx^N+\sum_{i=1}^3 d\mu_i^2+\mu_i^2\left(d\phi_i+\frac{2}{\sqrt{3}}A_\mu dx^\mu\right)^2
\label{10dmetric}
\ea
where $g_{MN}$ is the five-dimensional AdS metric, $\mu_i$'s are 2 independent the $S^5$ parameters with the constraint $\sum_{i=1}^{3}\mu_i^2=1$ and $\phi_i$'s are the remaining 3 angles of $S^5$.  The Kaluza-Klein reduction of this metric is shown \cite{Chamblin:1999tk,Cvetic:1999xp} to be a consistent truncation that generates the action (\ref{action}). To see that the Maxwell sector does not affect the axion sector it is sufficient to observe that the only place where the $U(1)_R$ field might show up in the action (\ref{axion}) is the ten-dimensional metric determinant $\sqrt{-G}$. However this is not possible because the gauge field can be thought of as a re-parametrization of the angles $\phi_i$ of $S^5$ under which $G$ is invariant. From the gauge theory perspective, there cannot be any gauge dependent term in the action; therefore $G$ has to be independent of $A_\mu$, which is in fact true. Also the NS-NS 2 form $B$ can be turned off consistently in the compactifiaction. As a result,  the axion field $C_0$ is a massless scalar in the five-dimensional Einstein-Maxwell theory.
\vskip0.3cm

 The associated Klein-Gordon equation for a generic metric of the form (\ref{metric}) is not analytically soluble, since we do not even know the metric explicitly. However it can be expressed in terms of the functions $U,V$ and $W$ that determine the metric at the linear order in frequency, which is the limit we are interested in. The dissipative processes that lead to an imaginary part in the retarded Green's function are associated with the in-falling boundary conditions for the dual scalar field near the black brane horizon. We shall study the Klein-Gordon equation in Eddington-Finkelstein coordinates where the in-falling boundary conditions near the horizon appear naturally. Let us define
\bea
t_*=t+\int_\infty^r\frac{r^\prime}{U(r^\prime)} ;
\ea
the metric (\ref{metric}) with this coordinate transformation becomes:
\bea
ds^2=-U(r)dt_*^2+2\,dr\,dt_*+e^{2V(r)}(dx_1^2+dx_2^2)+e^{2W(r)}dx_3^2
\label{metric_ef}
\ea
Note that the frequency remains unchanged: $\omega_*\sim i \del_{t_*}=i\del_t\sim\omega$. Furthermore, the wavefunctions are of the form $e^{-i\omega t_*+i \vec k\cdot \vec x} \phi_{\omega,\vec k} (r)$. The mode function $\phi_{\omega,\vec k} (r)$ should have regular near horizon behavior.  Near the horizon, $t_*\approx t+ \frac{1}{U^\prime(r_h)}\ln|r-r_h|$ up to constant, where $r=r_h$ is the horizon. The Hawking temperature for a general $U(r)$ is given by the relation:  $U^\prime(r_h)=4\,\pi\,T$.  Therefore the wavefunction has the expected infalling behavior $e^{-i\omega t}|r-r_h|^{-i \frac{ \omega}{4 \pi T}}$ which arises naturally in the Eddington-Finkelstein coordinates.  

The Klein-Gordon equation $\del_\mu(\sqrt{-g}\,g^{\mu\nu}\del_\nu\phi)=0$ with the metric (\ref{metric_ef}) reads:
\bea
\phi_{\omega,\vec k}^{\prime\prime}+\left(f(r)+\ln U(r)\right)^\prime\phi_{\omega,\vec k}^\prime-\frac{i \omega}{U(r)}(2\phi^\prime_{\omega,\vec k}+f^\prime(r) \phi_{\omega,\vec k})-\frac{1}{U(r)}(e^{-2V(r)}\vec k_\perp^2+ e^{-2W(r)}\vec k_\parallel^2)\phi_{\omega,\vec k}=0
\label{kg}
\ea
Here we defined $e^{f(r)} \equiv\sqrt{-g} = e^{2V(r)+W(r)} $ and $^\prime$ is short for $\del_r$. We now take $\vec k=0$ and solve (\ref{kg}) for small frequencies. To do this, we expand $\phi_\omega(r)$ to linear order in $\omega$ for $\omega<<1$\footnote{Here there is a slight abuse of notation. By writing $\omega<<1$ we actually mean $\frac{\omega}{4 \pi T}<<1$. But it is always possible to work with dimensionless frequency by rescaling $t_*$ and $r$ which we are doing by setting $U^\prime(1)=1$ (i.e. $4\pi T$=1). }.
\bea
\phi_{\omega,\vec k=0}=\phi_0(r)+i \omega \phi_1(r)+\ord(\omega^2)
\ea
The Klein-Gordon equation is now expanded in the same fashion as:
\bea
\phi_{0}^{\prime\prime}+(f+\ln U)^\prime\phi_0^\prime-i \omega\left(\frac{2}{U}\phi^\prime_{0}+\frac{f^\prime}{U} \phi_0-\phi_1^{\prime\prime}- (f+\ln U)^\prime\phi_1^\prime\right)=0
\label{kg_2}
\ea
which should be solved order by order in $\omega$. 

The $\ord(\omega^0)$ order equation has the solution:
\bea
\phi_0(r)=c_1+c_2\int^r\frac{dr^\prime}{\sqrt{-g(r^\prime)}U(r^\prime)}
\ea
The second term is divergent near the horizon since $U(r_h)=0$. In order $\phi_0(r)$ to be well behaved near the horizon, $c_2$ must vanish. We also set $c_1$=1. This choice translates into normalizing the homogeneous axion source in the boundary theory to unity in the AdS/CFT dictionary. 

After plugging $\phi_0(r)=1$ into (\ref{kg_2}), the $\ord(\omega^1)$ equation becomes:
\bea
\phi_1^{\prime\prime}+ (f+\ln U)^\prime\phi_1^\prime=\frac{f^\prime}{U}
\ea
which can be solved as:
\bea
\phi_1(r)=\int_{\infty}^r\frac{dr^\prime}{U(r^\prime)}(1-e^{f(r_h)-f(r^\prime)})= \int_{\infty}^r\frac{dr^\prime}{U(r^\prime)}\left(1-\frac{\sqrt{-g(r_h)}}{\sqrt{-g(r^\prime)}}\right)
\ea
Here the constant $\sqrt{-g(r_h)}$ is chosen to assure the regularity of the solution near the horizon. The other integration constant is determined by setting the limit of the integral such that $\phi_1$ vanishes near the boundary and does not effect the normalization of the source. 

\begin{figure}[h]
   \centering
   \includegraphics[scale=0.7]{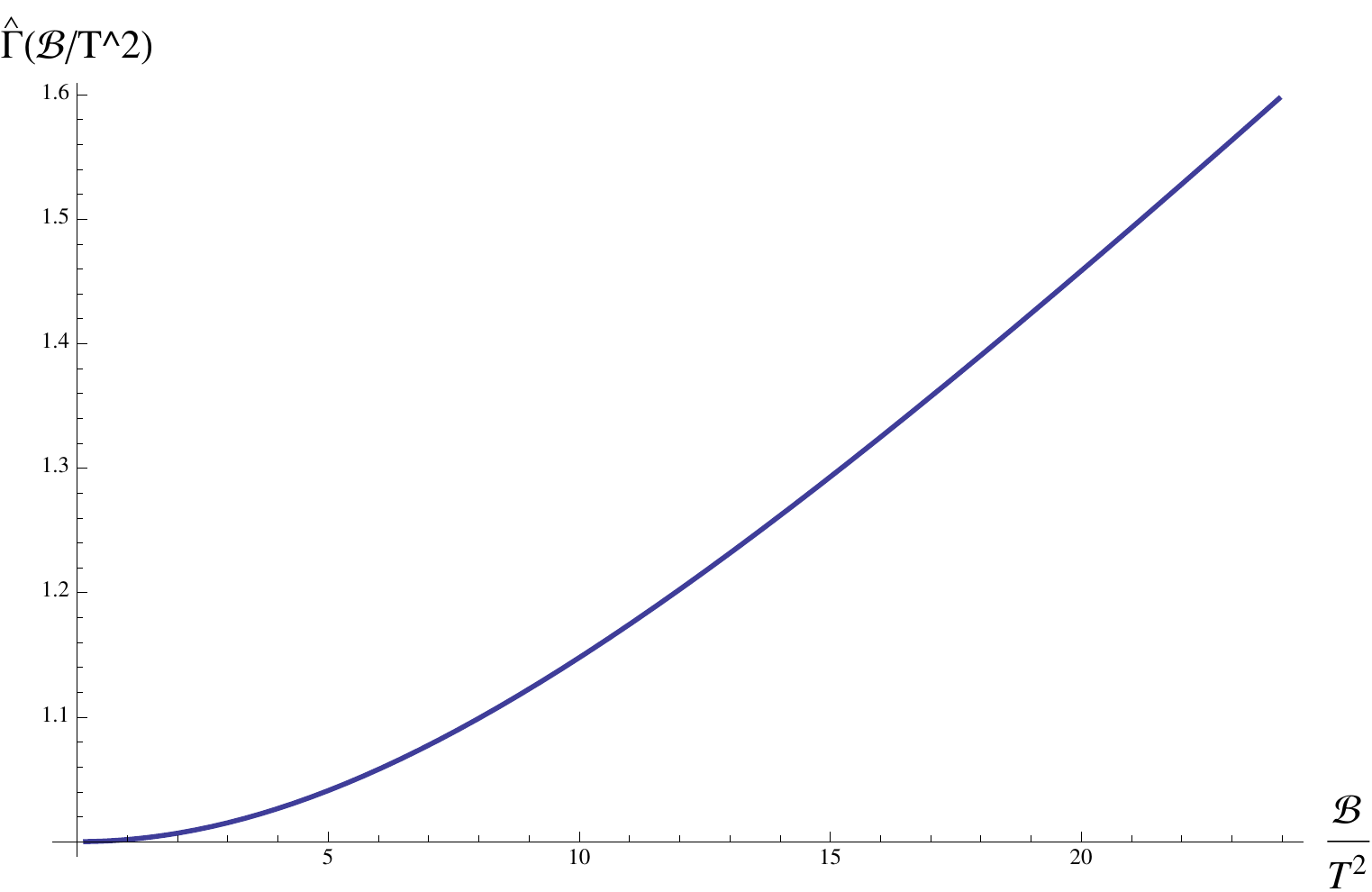} 
   \caption{The Chern-Simons diffusion rate $\hat \Gamma=\Gamma(\B,T)/\Gamma_0(T)$, normalized by the zero magnetic flux value $\Gamma_0=\frac{\lambda^2}{256\pi^3}T^4$ as a function of the dimensionless magnetic field $\B/T^2$. The diffusion rate is monotonously increasing with the magnetic field strength and has the asymptotic behavior $\Gamma\sim \B T^2$ for $\B>>T^2$ .}
   \label{fig1}
\end{figure}

As a result the solution of the Klein-Gordon equation (\ref{kg}) in the zero momentum, small frequency limit is:
\bea
\phi_{\omega,\vec k=0}&=&1+i\,\omega \int_{\infty}^r\frac{dr^\prime}{U(r^\prime)}\left(1-\frac{\sqrt{-g(r_h)}}{\sqrt{-g(r^\prime)}}\right)+\ord(\omega^2)\nn
&=&1+i\,\omega \int_{\infty}^r\frac{dr^\prime}{U(r^\prime)}\left(1-e^{2V(r_h)+W(r_h)-2V(r^\prime)-W(r^\prime)}\right)+\ord(\omega^2)
\label{phi}
\ea
We now expand (\ref{phi}) near the boundary. Recall that our metric converges to the $AdS_5$ metric $r^{-2}dr^2+r^2dx^\mu dx_\mu$ near the boundary $r\rightarrow\infty$. Therefore $\sqrt{-g}\rightarrow r^3$, $U(r)\rightarrow r^2$ and
\bea
\phi_{\omega,\vec k=0}(r)\approx1-\frac{i \omega}{r}+\frac{i \omega \sqrt{-g(r_h)}}{4 r^4}+\dots
\ea  
which reproduces the expected boundary behavior of a massless scalar. The first term is the non-normalizable mode that corresponds to the homogeneous axion source $\delta\Phi$ normalized to unity. The coefficient of the $r^{-4}$ term is: $\delta\langle\frac{1}{4} F^a_{\mu\nu} \tilde F_a ^{\mu\nu}\rangle$, that is the response of the system to the axion source $\delta \Phi$. Putting back all the constants it is found as:
\bea
\delta \left\langle \frac{1}{4} F^a_{\mu\nu} \tilde F_a ^{\mu\nu}\right\rangle=(2\Delta_{F\tilde F}-4) \kappa \frac{i \omega \sqrt{-g(r_h)}}{4}=i\,\kappa\,\omega \sqrt{-g(r_h)}
\ea 
Here $\Delta_{F\tilde F}=4$ is the scaling dimension of the topological charge operator. We can easily calculate the retarded Green's function in the zero momentum, small frequency limit through the linear response relation (Kubo's formula):
\bea
\delta\left\langle \frac{1}{4}F^a_{\mu\nu} \tilde F_a ^{\mu\nu}(\omega)\right\rangle =G^R(\omega)\, \delta \Phi
\ea
Thus we obtain $G^R$ in the leading linear frequency limit:
\bea
G^R(\omega,\vec k=0)=\frac{\delta\langle \frac{1}{4}F^a_{\mu\nu} \tilde F_a ^{\mu\nu}(\omega)\rangle}{\delta\Phi}= i\,\kappa\,\omega \sqrt{-g(r_h)}+\ord(\omega^2)
\label{retarded}
\ea
Using the properly scaled metric (\ref{metric2}) it reduces to:
 \bea
 G^R(\omega,\vec k=0)= \frac{i \omega \kappa}{v\sqrt{w}} +\ord(\omega^2)
 \ea
Recall that the diffusion rate is related to the retarded Green's function as:
\bea
\Gamma_{CS}&=&-\left(\frac{g^2}{8 \pi^2}\right)^2\, \lim_{\omega\rightarrow0}\frac{2 T}{\omega}\mc Im \left[G^R(\omega,\vec k=0)\right]\nn
&=&\left(\frac{g^2}{8 \pi^2}\right)^2\frac{N^2}{8\pi^2}\,\frac{2 T}{v\sqrt{w}}
\label{cs1}
\ea
In the last step we have used the holographic relation between the gravitational and gauge couplings: $\kappa=-\frac{1}{16 \pi G_5}=-\frac{N^2}{8 \pi^2}$ where $G_5$ is the five-dimensional Einstein constant. Recall that the functions $v$ and $w$ are obtained in the units $U(1)=0,\,U^\prime(1)=1$. In these units $T=1/4\pi$ and due to conformal symmetry we can choose our physical parameter to be the dimensionless magnetic field strength $\B/T^2$. We then express the diffusion rate in terms of this dimensionless ratio as follows:
\bea
\hat\Gamma(\B/T^2)=\frac{\Gamma(\B,T)}{\Gamma_0}=\frac{2^6}{v\sqrt{w}}\nn
\Gamma_0=\frac{(g^2N)^2}{256 \pi^3}T^4
\label{difrate}
\ea
This quantity measures the ratio of the diffusion rate in a medium with a magnetic flux to the one in a medium without a magnetic flux, $\Gamma_0$. From the plot (Fig.\ref{fig1}), we see that $\hat\Gamma(B/T^2)$ is a monotonously increasing function of the magnetic field strength. We now investigate the high and low temperature limits of the diffusion rate.

\section{High and low temperature limits}
\label{sec_limits}

\subsection{Weak magnetic field (high T) limit}

In the limit where $\B<<T^2$, the magnetic field can be treated as a perturbation to the $T^4$ scaling of the diffusion rate. Due to the R-charge symmetry of the gauge theory, we expect the perturbation series to start at  the order $\B^2$. Also we can argue that we do not expect the result to  depend on the direction of the magnetic field at the leading order and the analyticity of the perturbation series dictates the $\B^2$ dependence. It is also clear from the gravity side that the perturbative corrections to the Einstein equations (\ref{einstein}) will be at the order $\B^2$ since the magnetic field enters the equations in a gauge invariant way. Let us perturb the Einstein equations (\ref{einstein}) around the $AdS_5$ black brane solution:
\bea
U(r)=r^2-\frac{r_H^4}{r^2}+\B^2\,u_1(r)\nn
V(r)=\text{log}(r)+\B^2\,v_1(r)\nn
W(r)=\text{log}(r)+\B^2\,w_1(r)
\ea
where $r_H=\pi\,T$.
To extract the diffusion rate, we only need to know $\sqrt{-g}$ at the perturbed horizon. Let us call the perturbed horizon $r_P$, such that $U(r_P)=0$. By expanding to first order in $\B^2$ we get:
\bea
\sqrt{-g(r_P)}=e^{2V(r)+W(r)}=r_P^3\left(1+\B^2(2v_1(r_p)+w_1(r_p)\right)
\ea
From the linearized (in $\B^2$) Einstein equations we obtain the following relations:
\bea
u_1(r)=-\frac{2}{9\,r^2}(1 + 3 \log(r/r_p))\nn
2v_1^\prime(r)+w_1^\prime(r)=0
\ea
These corrections should vanish as $r\rightarrow\infty$ since we should have asymptotically $AdS_5$ behavior.  This condition sets the constant $2v_1(r)+w_1(r)=0$ leading to:
\bea
\sqrt{g(r_p)}=r_H^3\left(1+\frac{1}{6}\frac{\B^2}{r_H^4}\right)=\pi^3 T^3\,\left(1+\frac{1}{6\pi^4}\frac{\B^2}{T^4}\right)
\ea
As a result, to leading order in $\B^2$ we get:
\bea
\Gamma_{CS}= \frac{(g^2N)^2}{256 \pi^3}T^4\left(1+ \frac{1}{6\pi^4}\,\frac{\B^2}{T^4}+\ord(\frac{\B^4}{T^8})\right)
\ea 
Comparing with the exact numerical result (\ref{difrate}) we have found that this approximation differs from the exact result only by 2.5\% for $\B\sim10\ T^2$. This suggests that a proper expansion parameter is probably $\B^2/(\pi T)^4$.

 \subsection{Strong magnetic field (low T) limit and BTZ black hole}

For strong magnetic field $B>>T^2$, $v$ and $w$ have the limits
\bea
\lim_{\B\rightarrow\infty}\frac{1}{v\sqrt{w}}=\frac{B}{6 v (4\pi T)^2}=\frac{\B}{96\sqrt{3}\pi^2T^2}
\ea
In the last step we plugged in the physical magnetic field strength $\B=\sqrt{3}B/v$. Therefore in the presence of a strong magnetic field the diffusion rate is:
\bea
\Gamma(\B,T)=\frac{(g^2N)^2}{384 \sqrt{3} \pi^5}\B\, T^2\quad,\quad\B>>T^2
\label{strongbrate}
\ea
It is possible to obtain the same result in a way where the physics is more transparent. In \cite{D'Hoker:2009mm}, it was shown that there is another solution to Einstein-Maxwell configuration which is a product of a (2+1)-dimensional BTZ black hole in the $(r,t,x_3)$ directions and a flat surface in the transverse plane:
\bea
ds^2=-3(r^2-r_h^2)dt^2+\frac{dr^2}{3(r^2-r_h^2)}+3 r^2 dx_3^2+\frac{B}{\sqrt{3}}(dx_1^2+dx_2^2)
\label{btz}
\ea
The Hawking temperature is $T=3r_h/2\pi$. This is an appropriate description of the dual gauge theory in the strong magnetic field limit. When $\sqrt{B}>>T$, the fields occupy the lowest Landau levels and transitions to higher Landau orbits are suppressed by the factor $e^{-\sqrt{B}/T}$.  As a result the motion in the transverse plane is frozen and the only remaining degree of freedom is along the longitudinal direction $x_3$ where the motion is not affected by the magnetic field.  The overall factor of $B$ in front of the transverse plane corresponds to the density of the lowest Landau levels in the gauge theory.  The geometry (\ref{btz}) is therefore merely the holographic manifestation of the dimensional reduction of the system in the presence of a strong magnetic field. Using $\sqrt{-g}=r_hB$ and the expression (\ref{retarded}) for the retarded Green's function we obtain
 \bea
 G^R(\omega,\vec k=0)= i \omega \kappa B r_h = i \omega \kappa\,\frac{ 2 \pi T \B }{3\sqrt{3}} +\ord(\omega^2) .
 \ea
 
 Note that the expression (\ref{retarded}) is valid only for metrics that are asymptotically $AdS_5$ since the retarded Green's function is read off from the coefficient of $r^{-4}$ in the expansion of $\phi$ near the boundary. Here in the dimensionally reduced calculation, we assume that the magnetic field strength is very close to UV scale of the gauge theory so that the transition from BTZ to $AdS_5$ occurs close to the UV boundary and far away from the horizon in the dual picture. Therefore, the BTZ approximation is good near the horizon and we can use the BTZ metric for $\sqrt{-g(r_h)}$. 

 Plugging this result for $G^R$ into the expression for the diffusion rate (first line of (\ref{difrate})) leads to the correct strong magnetic field result (\ref{strongbrate}). This result can be interpreted as follows: $\Gamma_{CS}$ is a transport coefficient and its dynamics are governed by the IR sector of the theory. In the R -charged sector, the IR dynamics is dominated by the lowest Landau level fermions. The contribution of scalars to IR is suppressed exponentially since they have no zero modes as opposed to fermions. The holographic dimensional reduction mechanism explained above suggests that the strong interactions between the dimensionally reduced R-charged sector and  R-neutral non-Abelian gauge sector effectively induces a dimensional reduction for the gauge field in the IR, even though it does not see the magnetic field at the level of Lagrangian. Therefore the Chern-Simons number changing transitions occur in one spatial dimension which brings a factor of $T^2$ to the rate on dimensional grounds. This should be multiplied by the density of lowest Landau levels in transverse plane which scales as $\B/\pi$.

\section{Conclusion and Discussions}
\label{sec_conclusions}

In the strongly coupled $\mc N=4$ theory, the existence of an external $U(1)_R$ magnetic field increases the Chern-Simons diffusion rate in comparison to the zero magnetic field case. This modification is due to the strong interactions between the R-charged sector of the theory and the R-neutral Yang-Mills fields that is realized through the alteration of the dual gravity by the magnetic field. Furthermore in the strong magnetic field limit the Chern-Simons number transitions occur in the dimensionally reduced system and the rate thus scales as $B\times T^2$. The absence of the interaction between the $U(1)_R$ and Yang-Mills fields makes the $U(1)_R$ magnetic field a reasonable proxy for a ``real" U(1) magnetic field.   We thus expect a  similar behavior in a strongly coupled, non- SUSY gauge theory with a $U(1)$ magnetic field.
\vskip0.3cm
We would like to point out an interesting analogy between the electroweak sphaleron in a magnetic field and our result. The electroweak sphaleron in the Higgs phase (Klinkhamer-Manton sphaleron) \cite{Klinkhamer:1984di} is spherically symmetric for zero Weinberg angle. However a non-zero Weinberg angle reduces the symmetry into an axial one \cite{Klinkhamer:1990fi,Kunz:1992uh}. Furthermore, the deformed sphaleron develops a magnetic moment. Due to this dipole moment, an external magnetic field decreases the energy of the sphaleron which lowers the barrier between the topologically inequivalent vacua \cite{Comelli:1999gt,DeSimone:2011ek}. This puts some constraints on the electroweak baryogenesis scenario with a primordial magnetic field since a lower barrier means topological transitions are more frequent and they wash out any initial baryon number asymmetry. 
\vskip0.3cm
Our result suggests that  an external magnetic field raises the rate of topological transitions also in strongly coupled regime. In this case the energy of the sphaleron is altered by the magnetic field due to strong interactions. Furthermore,  the spherical symmetry is reduced to axial symmetry through dimensional reduction since there is a residual $O(2)$ symmetry in the transverse plane. These effects of the magnetic field are, of course, due to a different mechanism than the Kinkhamer- Manton sphaleron case above. However, it is noteworthy that even though the mechanisms are quite different, it seems a coupling to an extra $U(1)$ field affects the non-Abelian sector in a similar fashion for both cases. 
\vskip0.3cm
There are electroweak baryogenesis scenarios where there exist CP violating processes in the scalar (Higgs) sector \cite{baryo1,baryo2,baryo3,baryo4}. These processes can be described by an effective inhomogeneous axion field where chiral quarks are created locally on these inhomogeneities. These inhomogeneities correspond to the walls of the bubbles of broken phase nucleated inside the symmetric phase. As the created net chirality propagates into the symmetric phase it triggers the electroweak sphaleron transitions in the symmetric phase which ultimately creates the baryon number. However,  the strong sphaleron transitions damp the net chirality and therefore reduce the baryon number creation \cite{McLerran:1990de,Moore2}. Taking into account the existence of a primordial magnetic field, basing on our result we can conclude that it would further decrease the rate of baryon number generation if the relevant strong interactions are in the strongly coupled regime. Therefore, it seems that an existence of a primordial magnetic field works against the baryon number generation. 

Finally, let us pass from the big bang to the little bang: the heavy ion collisions. The background magnetic field created in these collisions is typically of the order $T\sim\sqrt{B}\sim m_{\pi}$ \cite{CME,bfield}. Our result suggests that a magnetic field of this strength has a negligible effect on the topological charge diffusion. To be more precise
\bea
 \frac{\Gamma(\B,T)-\Gamma(\B,0)}{\Gamma(\B,0)} \approx \frac{1}{6\pi^4}\approx0.0017 
 \ea
 for $\B=T^2$. Therefore the magnetic field changes the rate only by 0.17 \%. 
This means that the quantitative estimates of the chiral magnetic effect in heavy ion collisions can be safely done by using the sphaleron rate in the absence of magnetic flux.
 
 \section{Acknowledgements}
We are grateful to Ho-Ung Yee for a valuable advice, and to Gerald Dunne, Guy Moore and Larry McLerran for useful discussions. This work was supported by the US Department of Energy under grants  DE-AC02-98CH10886 and DE-FG-88ER41723.

\end{document}